\def\BibTeX{{\rm B\kern-.05em{\sc i\kern-.025em b}\kern-.08em
    T\kern-.1667em\lower.7ex\hbox{E}\kern-.125emX}}

\documentstyle[]{elsart}
\begin{document}
\bibliographystyle{unsrt}
\begin{frontmatter}
\title{Kinetic roughening and phase ordering in the two-component
growth model}
%
\author[FzU]{M. Kotrla\thanksref{e-Mirek}}
\thanks[e-Mirek]{Electronic address: kotrla@fzu.cz}
\author[IChPF]{, M. P\v{r}edota}
\author[FzU,CTS]{and F. Slanina}
\address[FzU]{
 Institute of Physics, 
 Academy of Sciences of the Czech Republic,\\
 Na~Slovance~2, 
 CZ-180~40~Praha~8, 
 Czech Republic
}
\address[IChPF]{ 
 Institute of Chemical Process Fundamentals,\\
 Academy of Sciences of the Czech Republic,\\
 CZ-165~02~Praha~6, 
 Czech Republic
}
\address[CTS]{
 Centre for Theoretical Study, 
 Jilsk\'a~1,
 CZ-11000~Praha, 
 Czech Republic 
}
\begin{abstract}
{\small
Interplay between kinetic roughening and phase ordering
is studied in a growth SOS model with
two kinds of particles and Ising-like interaction
by Monte Carlo simulations.
We found that,
for a sufficiently large coupling, 
growth is strongly affected by interaction
between species.
Surface roughness increases rapidly with coupling.
Scaling exponents
for kinetic roughening are enhanced with respect to 
homogeneous situation.
Phase ordering which leads to the lamellar structure persisting for a long time
is observed.
Surface profiles in strong coupling regime
have  a saw-tooth form,
with the correlation between the positions of local minima
and the domain boundaries.
}
\end{abstract}

\end{frontmatter}

\section{Introduction}
Growth by vapour deposition is a technologically important 
process for producing 
high quality materials. During last years much progress has been made in 
understanding of growth mechanisms on the microscopic level \cite{rev_sim}.
However, these results refer mainly to single-component homogeneous growth
and growth of binary 
materials is much less understood.
Essential feature in binary systems is phase ordering \cite{bray94}.
Recently there was interest in interplay between phase
ordering during growth
and kinetic roughening \cite{smith96,leonard97a,kotrla97c}.
Kinetic roughening is one of the important aspects of 
crystal growth. Due to stochastic fluctuations during a deposition process
the growing surface becomes rough and roughness typically increases
with time as a power with a characteristic exponent $\beta$.
This is important from the practical point of view,
besides kinetic roughening also attracted a lot of interest
in the context of non-equilibrium statistical 
mechanics \cite{rev_kr}.

In this paper we discuss the interplay between compositional 
ordering and kinetic roughening in a simple model for growth 
of binary alloys which we recently formulated \cite{kotrla97c}.

\section{Model}
Our model is based on the single-step solid-on-solid geometry.
We confine ourselves here to one-dimensional substrate
with the coordinate $i$.
The surface is described
by a single-valued function $h_i(t)$
with the additional constraint $|h_i - h_j|=1$ for neighbouring
sites $i$ and $j$.
A new particle can be added only at a growth site
corresponding to a local minimum in the surface height profile.
We consider two types of particles, which
we distinguish by a variable
$\sigma$ taking the value $+1$ or $-1$.
Growth rules are controlled by a change of energy of
alloy after deposition of a new particle. The energy is 
given by the Ising-like interaction.
The probability of deposition of a new particle
of type $\sigma $ to a growth site $i$
is proportional to $\exp (-\Delta E(i,\sigma)/k_BT)$;
$k_B$ is Boltzmann's constant and $T$ is temperature.
The change of energy is
$\Delta E(i,\sigma )/k_BT=-K\sigma \left[ \sigma (i-1)+\sigma (i)+\sigma
(i+1)\right] -H\sigma $.
Here $K$ is a dimensionless coupling strength
and $H$ is an external field.
We call this model a two-component single-step (TCSS) model.
We consider here only the case of zero external field.

The surface roughness is described by the surface width
$w(t,L)=\langle\sqrt{\overline{h^2}-\overline{h}^2}\rangle$,
where $t$ is the time, $L$ is the linear size
and the bar denotes a spatial average,
$\langle ... \rangle$ a statistical average.
Ordering is studied by the time evolution
of the average domain size along the surface $D(t)$.

%
\begin{figure}[hb]
\centering
\vspace*{61mm}
\includegraphics{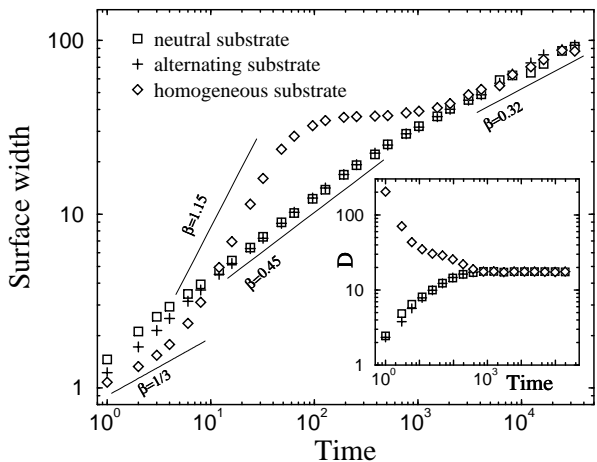}
\caption{{\small Surface width $w$ {\it vs.} time for
coupling strength $K=1.0$, and three different 
types of the substrate of size $L=10000$. 
Data are averaged over 30 independent runs.
The values of slope, $\beta =0.45$ and $\beta =0.32$, were obtained by fit to
points for the alternating substrate, the straight line with slope $1/3$ is only
guide for eye.}}
\label{fig:rough}
\end{figure}

\section{Results}
We performed extensive simulations for various
values of $K$.
Evolution of the surface profile is affected
by composition of the surface, in particular it depends
on composition of the initial flat substrate.
We considered three different types:
i) a neutral substrate, i.e. without any interaction with deposited particles,
ii) an alternating substrate, with alternating types of particles and
iii) a homogeneous substrate composed of one type of particles.
\begin{figure}[hb]
\centering
\vspace*{120mm}
\includegraphics{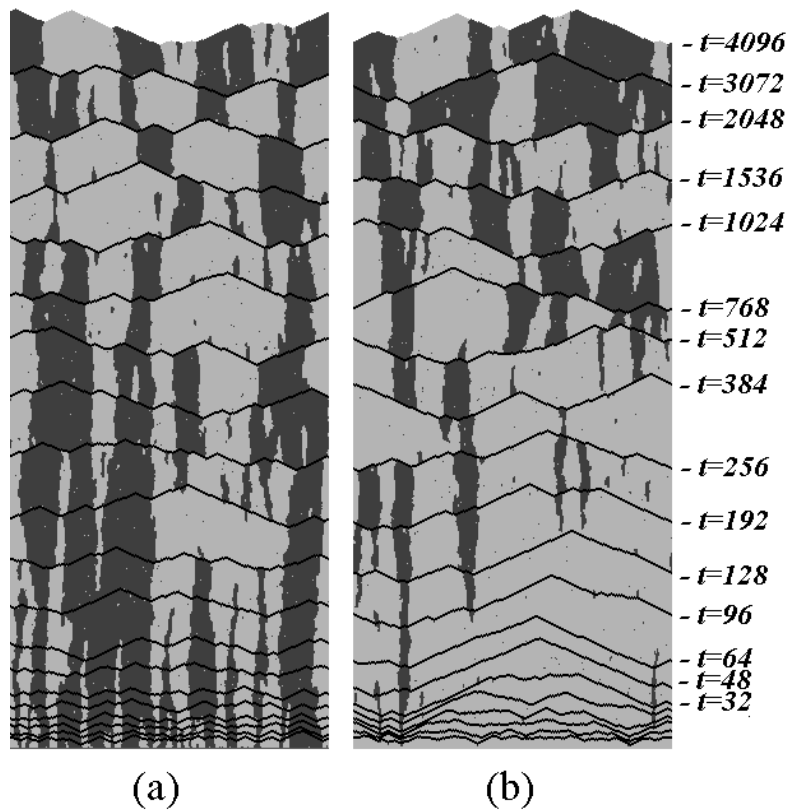}
\caption{Examples of evolution of surface profiles for
$K=1.0$, and two different substrates:
(a) the neutral substrate and (b) the homogeneous substrate.
Surface profiles at various times increasing as powers
are indicated by black lines,
only part of the grown material close to the surface is shown for given time.
Dark and light grey correspond to different types of particles.
System size is $L=250$.}
\label{fig:conf}
\end{figure}

Evolution on the neutral and the alternating substrate is almost the same.
In  Fig. \ref{fig:rough}
we can see  that the effective exponent ($\beta_{\rm eff} =0.45$)
is at first larger than the exponent for homogeneous growth,
$\beta=1/3$, which corresponds to the so-called Kardar-Parisi-Zhang (KPZ)
class \cite{kardar86}. However, after a certain time 
$t_{\rm cross} \approx 10^3$
the effective exponent crosses over to the value close to $1/3$
($\beta_{\rm eff} =0.32$).
This behaviour can be explained by evolution of compositional ordering.
In the inset of Fig. \ref{fig:rough} we present time dependence of 
the average domain size for three substrate types considered. 
We can see that for all of them the domain size saturates at time close
to $t_{\rm cross}$.
Hence, ordering of the surface leads to enhanced increase of the roughness,
but when ordering stops the surface roughness
increases further with the same exponent as in one-component single-step
growth model.

In the case of the homogeneous substrate
the behaviour of the surface width changes 
dramatically (cf.  Fig. \ref{fig:rough}).
The general features 
are the following.
At first the width starts to grow with the exponent close to $1/3$ (this
regime is extended over several decades of deposition time for 
very strong coupling).
It is caused by the fact that practically no domains with minority
type of particles are formed
because probability to nucleate these domains
is rather small; growth is the same as in the pure single-step model.
Then the slope crosses over to a very large value ($\beta_{\rm eff} =1.15$).
This unexpected effect is related to growth of trapezoidal 
features with the length given by positions of the minority 
type domains, which start to appear but often again quickly disappear.
The rapid increase of roughness
ceases when relatively stable minority type domains 
are nucleated. The further increase of roughness is very slow
up to the time when the average domain size saturates
and growth continues with the KPZ exponent $\beta =1/3$.
We illustrate the difference in evolution on the neutral and on the homogeneous
substrate in Fig. \ref{fig:conf}. 
We can see that for the homogeneous substrate
dark domains appear very slowly.

\begin{figure}[thb]
\centering
\vspace*{100mm}
\includegraphics{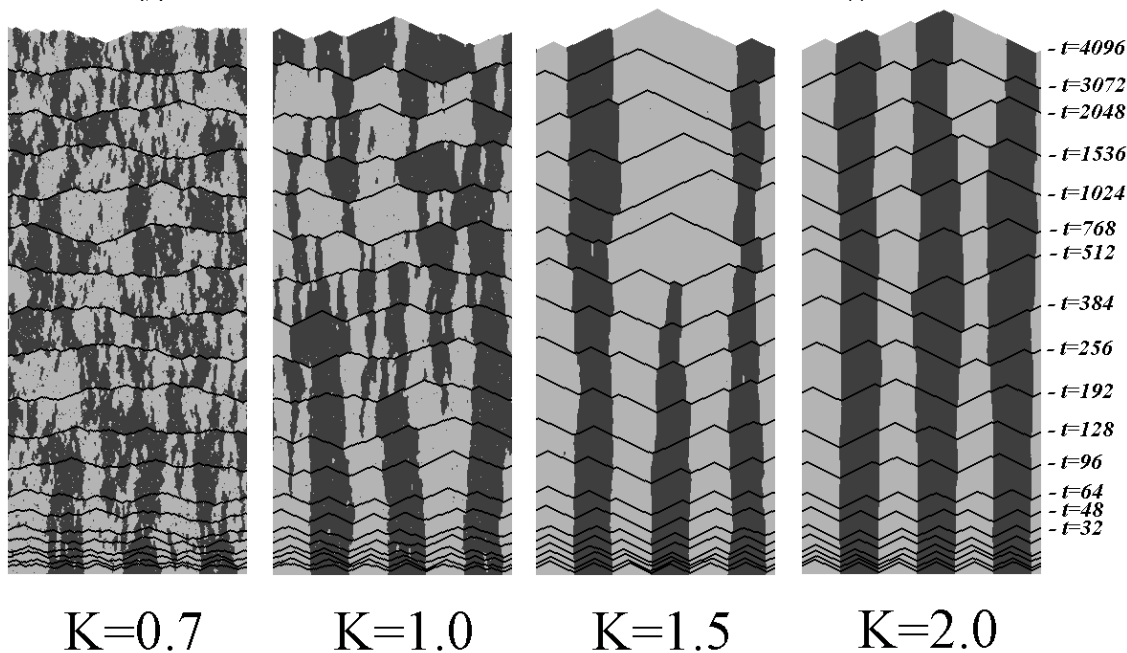}
\caption{Examples of evolution of surface profiles 
on the substrate with alternating domains of different particle types
for several coupling strengths $K$.
Meaning of colours and other conditions are the same as in Fig. 2.}
\label{fig:lamel}
\end{figure}

Let us now turn to the dependence on the coupling $K$ in the case 
of the neutral substrate.
For small $K$, there is no ordering in the growth direction.
Small domains appear and shrink again, but for a larger $K$ the 
clear lamellar structure exists.
We illustrate this in Fig. \ref{fig:lamel} 
which shows evolution
of configurations on a substrate with periodic domains
imitating the lateral superlattice.
We can see that for $K=2$ lamellae persist for long time.
Notice the correlations between the positions of local minima
of surface profile and domain boundaries.
We observed that this is a typical case for stronger coupling
independently on the type of the substrate.
We have found that the roughness increases rapidly with the coupling.
The effective exponent $\beta_{\rm eff}$
for initial growth increases with $K$
and it seems to approach the value $1/2$;
for $K=2$ we have found $\beta =0.52 \pm 0.02$.
The crossover time for change of  $\beta_{\rm eff}$ to the KPZ exponent
is also increasing rapidly with $K$ ($t_{\rm cross} \propto e^{\gamma K}$
with $\gamma \approx 3.5$).

\section{Conclusion}
We demonstrated that phase ordering leads to faster kinetic roughening
than in the homogeneous case.
However, after some time
this behaviour crosses over to the ordinary behaviour for homogeneous growth.
But the crossover time increases rapidly
with interaction between particles,
so practically only the regime with enhanced roughness can be seen
for stronger coupling.
We also illustrated the strong sensitivity of evolution
on the initial composition of the substrate.

~~~~~{\bf Acknowledgement}\\
This work was
supported by grant No. A 1010513 of the GA AV \v{C}R.

\bibliography{../../my_bib/journals,../../my_bib/md,../../my_bib/exper,../../my_bib/simgr,../../my_bib/kr,../../my_bib/gen}

\newpage

\end{document}